# Finite-size scaling in Mott Metal-Insulater Transition on half-filled non-partite lattice

J. X. Wang and Sabre Kais

*Department of Chemistry, Purdue University West Lafayette, IN 47907*

By applying a multi-stage real-space renormalization group procedure to Hubbard model, we examine the finite-size scaling in Mott metal-insulator transition(MIT) on a non-partite lattice. It is found that there exists a critic point U/t=12.5 for the MIT and the corresponding critical exponent for correlation length $\nu = 1$ and the dynamic exponent $z = 0.91$ are obtained. On the critical point, the charge gap scales with the system size as $\triangle_g \sim 1/L^{0.91}$.

Ever since the experimental observations of the metallic behavior in 2-dimensional electron gas with high-mobility Si metal-oxide-semiconductor field effect (Si MOSEFT′s) [1], there has been an intensive interest in the investigations of electronic transport properties in such systems [2, 3]. The significance of this discovery is that it has taken us to a new regime where the electron correlations begin to play a very important role. For example, in the disordered systems, when the electron interactions $E_{ee}$ are much stronger than the Fermi energy $E_f$, namely, $E_{ee}/E_f$ lies in the range 5-50, they can show metallic properties. this is in quite contradiction to the conventional prevailing non-interacting-electron scaling theory, which states that for 2D systems, any disorder will localize all states and there should be no metal-insulator transition(MIT) [4]. Hence the early experiments have stimulated a spate of new experimetal results and MIT have been found in various 2D systems, such as p-type SiGe structures [5], p-type GaAs/AlGaAs heterostructures [6], n-type AlAs heterostructures [7] annd n-type GaAs/AlGaAs heterostructures [8] etc., but there have been no satisfactory theoretical explanations of this phenomena, to date. In theory, it is clear now that MIT in 2D systems can be driven by the disorders, the electron filling or the electron correlations with fixed density. In the domain of disordered system with no or weak electron interactions, where Anderson MIT [9] dominates, the scaling theory from Abraham [4] holds well and have been studied in every detail. When disorder effects is comparable to the influence of the electron interactions, the theory becomes very subtle and the nature of MIT is still a open question, which is actually becoming a central problem in condensed matter phyaics [10, 11]. If we go a little further, one is easy to ask: what will happen in the strong-coupling regime, namely the systems have very strong electron interactions with little or no disorder? Actually, this is the regime of the famous Mott MIT [12]. Although there have been much work in this direction [13], not much study is carried out upon the correlation induced or fixed density MIT in 2D systems from the viewpoint of the finite-size scaling(FSS) analysis [15]. As we know, FSS for Anderson MIT has a noteworthy history and a lot of work has been done on it [16, 17]. Hence, to carry out FSS upon Mott MIT is the main motivation for our following investigations.

The model we use is the Hubbard model [18]. In order to capture the physics mainly from electron interactions, we don't introduce disorders here. The Hamiltonian is written as,

$$H = -t \sum_{<i,j>,\sigma} [c_{i\sigma}^+ c_{j\sigma} + H.c.] + U \sum_i n_{i\uparrow} n_{i\downarrow}$$
$$-\mu \sum_i (n_{i\uparrow} + n_{i\downarrow}), \quad (1)$$

where $t$ is the nearest-neighbor hopping term, $U$ is the local repulsive interaction and $\mu$ is the chemical potential. $c_{i\sigma}^+(c_{i\sigma})$ creates(annihilates) an electron with spin $\sigma$ in a Wannier orbital located at site $i$; the corresponding number operator is $n_{i\sigma} = c_{i\sigma}^+ c_{i\sigma}$ and $<>$ denotes the nearest-neighbor pairs. H.c. denotes the Hermitian conjugate.

We will only consider the half-filled system since the corresponding electron interactions are most prominent in this case, which leads to $\mu = U/2$. Hence, Eq.(1) can be rewritten as,

$$H = -t \sum_{<i,j>,\sigma} [c_{i\sigma}^+ c_{j\sigma} + H.c.]$$
$$+U \sum_i (\frac{1}{2} - n_{i\uparrow})(\frac{1}{2} - n_{i\downarrow}) + K \sum_i I_i \quad (2)$$

with $K = -\frac{U}{4}$ and $I_i$ the unit operator. As to the lattice structure, we use the non-partite triangular lattice, for MIT emergers at finite $U = U_c$. It is well known that MIT on square lattice can only take place at $U = 0$ due to the perfect nesting of the Fermi suface. If we don′t study the exotice case with $U < 0$, which is possible in a strongly polarizable medium, the square lattice is not an optimal option for our purpose. Further more, the magnetic properties are not the interest of this paper, hence the only physical quantity we are concerned will be the charge gap, $\triangle_g$, which is defined as

$$\triangle_g = E(N_e - 1) + E(N_e + 1) - 2E(N_e) \quad (3)$$

where $E(N_e)$ denotes the lowest energy for a $N_e$−electron system. In our case, $N_e$ is equal to the site number $N_s$

of the lattice. This quantity is the discretized second derivative of the ground state energy with respect to the number of particles, i.e. the inverse compressibility.

Conventionally, Monte Carlo and exact diagonalization methods are two most used ways to carry out finite-size scaling analysis. But the first one is not suitable to study zero-temperature phase transition and the second one involves too intensive calculations, especially for large-size systems. In our work, we develop a multi-stage block renormalization group(RG) method to attack this problem.

The essence of real-space RG method is to map the original Hamiltonian to a new Hamiltonian with much fewer freedoms which keeps the physical quantities we are interested in unchanged [19]. The map can be iterated until the final Hamiltonian can be easily handled. The crucial step in this method is how to related the parameters between the old and the new Hamiltonians. This can be realized by dividing the original lattice into blocks and then build the new Hamiltonian upon blocks, namely regard each block to be an effective site.

Fig.1 shows schematically the hexagonal block structure we use in our calculations and the coupling between blocks. For each block, we solve it numerically in the subspaces of 6 electrons with 3 spin up and 3 spin down, 7 electrons with 4 spin up and 3 spin down, 7 electrons with 3 spin up and 4 spin down, and 8 electrons with 4 spin up and 4 spin down. In all the subspaces, we keep the lowest energy state, which is also required to belong to the same irreducible representations of $C_{6v}$ symmetry group. The kept states will then be taken as the 4 states for an effective site. If denoting the energies corresponding to the first two states by $E_1, E_2$, after some intensive calculations, we can obtain the new Hamiltonian on the effective lattice, which has the same structure as the original Hamiltonian, i.e.,

$$H' = -t' \sum_{<i,j>,\sigma} [c'^+_{i\sigma} c'_{j\sigma} + H.c.]$$
$$+ U' \sum_i (\frac{1}{2} - n'_{i\uparrow})(\frac{1}{2} - n'_{i\downarrow}) + K' \sum_i I_i, \quad (4)$$

where the prime ' denotes the operator action upon the block states and

$$t' = \nu \lambda^2 t, \quad (5)$$
$$U' = 2(E_1 - E_3), \quad (6)$$
$$K' = (E_1 + E_3)/2. \quad (7)$$

The above equations are the so-called RG flow equations. Usually, they are iterated until we get the fixed point. The charge gap for an infinite lattice can then be written as,

$$\Delta_g = \lim_{n \to \infty} U^{(n)}. \quad (8)$$

Because of the implicit functional in Eqs.(5-7), it is very difficult to obtain any other useful information except the critical transition point $U_c$ from the above procedures. In the following, we will handle these procedures in another way, namely, instead of letting RG flow to infinity for a fixed intial parameters$(U, k, t)$, we can stop the RG flow at some stage. Thus the energy gap got from Eq.(3) will correspond to a system of fixed size. For example, if we stop the RG flow at the first iteration, then obtained t' and U' will be for hexagonal block mapped from a system of $7^2$ sites. Because we can solve a hexagonal block Hamiltonian exactly, then the energy gap for a system of 49 sites can be got easily.

In this way, we can study the variations of $\triangle_g$ against the system size of $7^2, 7^3, 7^4, 7^5, 7^6$.....We call this procedure to be multi-stage real-space RG method, which is well adapted to start the finite-size scaling analysis.

In Fig.2(a), the size dependence of relationship between $\triangle_g$ and $U$ in the unit of $t$ is presented. But from this figure, it is somewhat not so easy to decide the transition point for $\triangle_g$ to reach zero as $U$ is decreased from big values. To explicitly display the critical phenomenon, we give Fig. 2(b) by scaling $\triangle_g$ with respect to $N$ at first. Here $N = N_e = N_s$. Now it is very easy to pin the critical value of $(U/t)_c$ at the crossing point of all the curves corresponding to different system size, which is found to be 12.5. The same value is obtained by letting RG equations flow to infinity as done in Ref.(16), which should be expected. But this kind of scaling gives us more. In Fig.2(c), all the data collapse to one curve once we carry out a second step of scaling with $U/t - (U/t)_c$ by $N$. It is an obvious evidence for the occurrence of a quantum phase transition with $U$ to be the tuning parameter. Hence we can write down the following equations,

$$\triangle_g N^{0.405} = f\left[q N^{0.5}\right], \quad (9)$$

where $f(x)$ is a universal functions independent of the size and $q = U/t - (U/t)_c$ measures the distance of the electron correlations from its critical value for MIT. By using $N = L^2$ for 2D systems, the above equation can be rewritten as,

$$\triangle_g = L^{-0.91} f[qL], \quad (10)$$

from which we can get two scaling relationship for the charge gap. One is the finite-size scaling at the transition point, i.e. when $q = 0$, $\triangle_g \sim 1/L^{0.91}$. As shown in Ref.[14], in Anderson MIT, when the electron correlation energy dominate the Fermi energy, the average inverse compressibility $(= \Delta_g)$ exhibits a scaling as $1/L$ with respect to the system size. Here it shows a slower decay as $L$ increases. The other one is the bulk scaling scaling around the transition point for infinite systems, $\triangle_g \sim q^{0.91}$. According to the scaling analysis of the Gutzwiller solution for the Mott MIT in the Hubbard model [15], $\triangle_g \sim q^{0.5}$. Since the Gutzwiller solution is a mean field approximation and the upper critical dimension for it to give a correct description of this critical phenomenon is $d_c = 3$, it is understandable that our above 2D results cannot be merged into the mean-field theory.

By introducing a critical exponent $y_\triangle$ for $\triangle_g$, from the one-parameter scaling theory, we can have

$$\triangle_g = q^{y_\triangle} f\left(\frac{L}{\xi}\right), \qquad (11)$$

in which $\xi = q^{-\upsilon}$ is the correlation length with $\upsilon$ to be the corresponding critical exponent and $L$ denotes the system size. By using $N = L^2$ for 2D systems, the above equation can be rewritten as,

$$\triangle_g = N^{-(y_\triangle/2\upsilon)} f\left(qN^{1/2\upsilon}\right). \qquad (12)$$

Comparing Eq.(9) and Eq.(10), we can easily get

$$y_\triangle = 0.91, \upsilon = 1 \qquad (13)$$

By relating $y_\triangle$ to the dynamic exponent $z$ with $y_\triangle = z\nu$, we can have $z \approx 91$. It should be meaningful to compare the obtained results here with those from other kinds of MIT in 2D systems.

For filling-control or density-driven MIT, there are two-types of universality classes. One is characterized by $z = 1/\nu = 2$, which is the case for all 1D systems as well as of several transitions at higher dimensions, such as transitions between insulator with diagonal order of components and metal with diagonal order components and small Fermi volume. Another one is characterized with $z = 1/\nu = 4$. Numerical calculations have shown that Hubbard model on a square lattice is an example of this class. The large dynamic exponent is associated with suppression of coherence, which is associated with strong incoherent scattering of charge by a large degeneracy of component excitations.

For Anderson MIT, most of the analytical, numerical and experimental researches have produced $\upsilon > 1$, such as $\upsilon = 1.35$ [21],1.54 [22],1.62 [23],....

Our work leads to $z = 0.91$ and $\nu = 1$, which might imply that Mott MIT does belong to any universality class mentioned above. But because of the approximations involved in real-space RG method and the sensitivity of the critical exponents upon the used approaches, more work, especially analytical work is much needed to reveal the implications and the underlying mechanism for $z = 0.91$ and $\upsilon = 1$. Since we lack in the results from other approaches for comparisons, $z$ and $\nu$ are thus far from decided. Our work tends to show $z \approx \nu \approx 1$. The research aiming to check this guess is still in progress.

In summary, by using a multi-stage real-space renormalization group method, we show that the finite-size scaling can be applied in Mott MIT. And the dynamic and correlation length critical exponents are found to be $z = 0.91$ and $\upsilon = 1$, respectively. At the transition point, the charge gap scales with size as $\triangle_g \sim 1/L^{0.91}$. Around the point, $\triangle_g \sim q^{0.91}$..

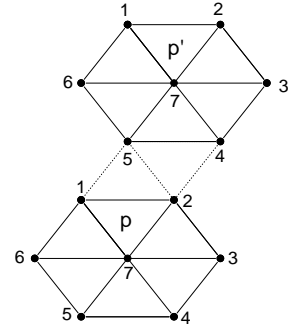

FIG. 1: Schematic diagram of the triangular lattice with hexagonal blocks. Only two neighboring blocks $p$ and $p'$ are drawn here. The dotted lines represent the interblock interactions and solid line intrablock ones.

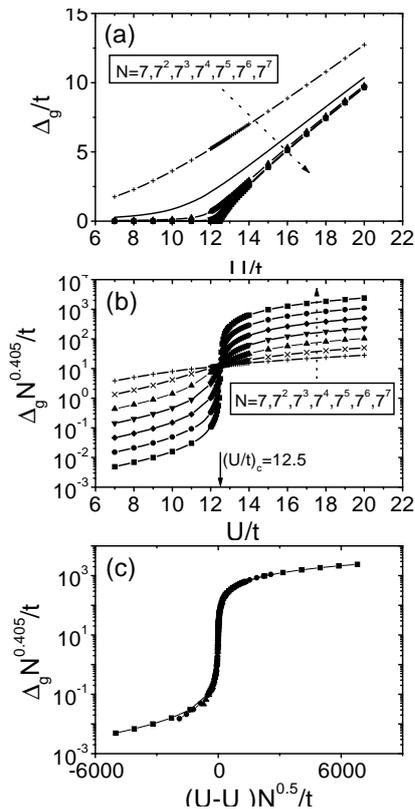

FIG. 2: Variations of the charge gap $\Delta_g$ against the on-site electron interation $U$ for different system size, i.e. the number of sites: $7$(cross +), $7^2$(cross ×), $7^3$(up triangle), $7^4$(down triangle), $7^5$(diamond), $7^6$(circle), $7^7$(square). More points are calculated around the transition point. In (a), no scaling is utilized. In (b), the charge gap is scaled by $1/N_0.405$ to display clearly the phase transition. In (C), all the data are collapsed onto one curve by scaling both axis with respect to $N$.